\documentclass[aps,prd,twocolumn,superscriptaddress,notitlepage,nofootinbib,preprintnumbers]{revtex4-2}

\usepackage{epsfig}
\usepackage{multirow}
\usepackage{slashed}
\usepackage{amsmath}
\usepackage{physics}
\usepackage{qcircuit}
\usepackage{braket}
\usepackage{graphicx}
\usepackage{subfigure}
\usepackage{color}
\usepackage{epstopdf}
\usepackage[inline]{enumitem}
\usepackage{tabularx}
\usepackage{booktabs}
 
\def\bea#1\eea{\begin{align}#1\end{align}}
\newcommand{\nn}{\nonumber\\}
\newcommand{\bef}{\begin{figure}[!htp]}
\newcommand{\eef}{\end{figure}}

\usepackage{xcolor}
\usepackage[normalem]{ulem}

\usepackage[colorlinks,
linkcolor = blue,
anchorcolor=blue,
citecolor=blue]{hyperref}

\begin{document}
\preprint{RIKEN-iTHEMS-Report-26}

\title{Hadronic tensor in lattice gauge theories by quantum computing}

\date{\today}

\author{Dairui Zou}
\affiliation{State Key Laboratory of Nuclear Physics and Technology, Institute of Quantum Matter,
South China Normal University, Guangzhou 510006, China}
\affiliation{Key Laboratory of Atomic and Subatomic Structure and Quantum Control (MOE), Guangdong-Hong Kong Joint Laboratory of Quantum Matter, Guangzhou 510006, China}
\affiliation{Guangdong Basic Research Center of Excellence for Structure and Fundamental Interactions of Matter,
Guangdong Provincial Key Laboratory of Nuclear Science, Guangzhou 510006, China}

\author{Tianyin Li}
\email{tianyin.li@riken.jp}
\affiliation{RIKEN Center for Interdisciplinary Theoretical and Mathematical Sciences (iTHEMS), RIKEN, Wako 351-0198, Japan}

\author{Jian Liang}
\email{jianliang@scnu.edu.cn}
\affiliation{State Key Laboratory of Nuclear Physics and Technology, Institute of Quantum Matter,
South China Normal University, Guangzhou 510006, China}
\affiliation{Guangdong Basic Research Center of Excellence for Structure and Fundamental Interactions of Matter,
Guangdong Provincial Key Laboratory of Nuclear Science, Guangzhou 510006, China}

\author{Enke Wang}
\email{wangek@scnu.edu.cn}
\affiliation{State Key Laboratory of Nuclear Physics and Technology, Institute of Quantum Matter,
South China Normal University, Guangzhou 510006, China}
\affiliation{Guangdong Basic Research Center of Excellence for Structure and Fundamental Interactions of Matter,
Guangdong Provincial Key Laboratory of Nuclear Science, Guangzhou 510006, China}

\author{Hongxi Xing}
\email{hxing@m.scnu.edu.cn}
\affiliation{State Key Laboratory of Nuclear Physics and Technology, Institute of Quantum Matter,
South China Normal University, Guangzhou 510006, China}
\affiliation{Guangdong Basic Research Center of Excellence for Structure and Fundamental Interactions of Matter,
Guangdong Provincial Key Laboratory of Nuclear Science, Guangzhou 510006, China}
\affiliation{Southern Center for Nuclear-Science Theory (SCNT), Institute of Modern Physics,
Chinese Academy of Sciences, Huizhou 516000, China}

\collaboration{QuNu Collaboration}

\date{\today}         

\begin{abstract}
The hadronic tensor encodes crucial information regarding the internal structure of hadrons, reflecting the non-perturbative features of quantum chromodynamics (QCD). In this work, we directly compute the hadronic tensor within (1+1)-dimensional $\rm U(1)$ and $\rm SU(2)$ gauge theories by evaluating real-time current-current correlation functions. Utilizing quantum algorithms executed on classical hardware, we demonstrate that the hadron form factors for both meson and baryon states can be reliably extracted from the hadronic tensor. Our methodology is validated by strong agreement with both direct calculation and exact diagonalization of the form factors.
\end{abstract}

\maketitle

\section{Introduction}\label{sec:one}

\begin{figure}[htbp]
	\centering
	\includegraphics[width=0.45\textwidth]{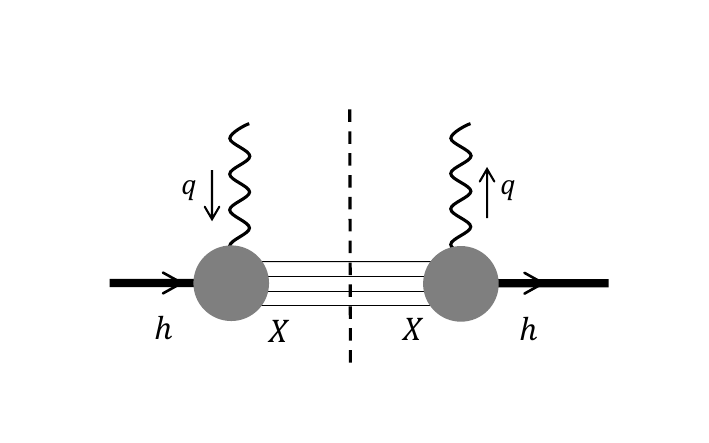}
	\caption{Schematic representation of the hadronic tensor in inclusive lepton-hadron scattering. The insertion of two electromagnetic currents sums over all allowed intermediate hadronic states $X$.}
	\label{fig:hadronic_part}
\end{figure}

Exploring the internal structure of hadrons remains a central goal in high-energy and nuclear physics. Among the various probes available, lepton-hadron scattering serves as a primary experimental avenue, having pioneered our understanding of hadron structure since the 1970s. In these scattering processes, the cross section is governed by the contraction of the leptonic tensor $L_{\mu\nu}$ and the hadronic tensor $W^{\mu\nu}$, schematically written as $d\sigma \propto L_{\mu\nu}W^{\mu\nu}$. For inclusive lepton-hadron scattering, the non-perturbative hadronic information is entirely encoded within this hadronic tensor. In $D$-dimensional space-time, it is defined as~\cite{Collins:2011zzd}
\bea \label{eq:hadronic_tensor}
W^{\mu\nu}(q) &= \int d^D z\,e^{iqz} \bra{h} J^{\mu}(z) J^\nu(0) \ket{h} \,,
\eea
where $\ket{h}$ denotes the initial hadron state and $q$ is the momentum of the virtual photon. Considering the electromagnetic interaction between the incident lepton and the hadron, the electromagnetic current operator is given by $J^\mu=e_f\bar{\psi}\gamma^\mu\psi$, where $e_f$ is the electric charge of the fermion $f$. As schematically illustrated for one-photon exchange in Fig.~\ref{fig:hadronic_part}, the two current insertions create and annihilate all allowed intermediate hadronic states $X$. The resulting current-current correlation function consequently captures the relevant non-perturbative QCD dynamics.

The physical information extracted from the hadronic tensor strongly depends on the kinematic scale of the momentum transfer, $Q^2\equiv -q^2$. When $Q^2 \gg \Lambda_{\rm QCD}^2$, the process enters the deep inelastic scattering (DIS) regime. In this high-$Q^2$ limit, the hadronic tensor factorizes into a convolution of perturbatively calculable hard coefficients and non-perturbative parton distribution functions (PDFs), schematically given by $W^{\mu\nu}(q) \sim \sum_a c^a_i \otimes f_{a/h}(x)$. Conversely, in the low-to-intermediate $Q^2$ region, where $Q^2 \sim m_h^2$ ($m_h$ being the hadron mass), the available momentum transfer is restricted. Here, the hadronic tensor predominantly probes hadron form factors, which characterize the spatial distribution of the hadron's constituents and its low-energy excitation dynamics.

Owing to the characteristic hadronic scales involved, the hadronic tensor is inherently non-perturbative and cannot be computed using perturbative QCD alone. While traditional lattice QCD in Euclidean space provides a powerful first principle framework, with extractions of the hadronic tensor from four-point correlation functions already successfully demonstrated~\cite{Liang:2019frk}, it faces a severe challenge. The physical Minkowski space response must be reconstructed from Euclidean-time data, an inverse problem that is notoriously ill-posed numerically. Quantum simulation offers a natural alternative through the Hamiltonian framework, where real-time dynamics can be simulated directly at the quantum state level. This capability has driven a surge of quantum-computing studies targeting real-time non-perturbative quantities~\cite{Fang:2024ple,Zhang:2020uqo,Bauer:2022hpo,Bauer:2023qgm}, such as scattering amplitudes~\cite{Jordan:2011ci, Li:2023kex}, PDFs~\cite{Lamm:2019uyc,Li:2021kcs, Kang:2025xpz, Banuls:2025wiq}, light-cone distribution amplitudes~\cite{Li:2022lyt}, fragmentation functions~\cite{Li:2024nod}, and parton showers~\cite{Barata:2025hgx}.

Quantum algorithms for parton physics and hadronic tensor were initially proposed in Ref.~\cite{Lamm:2019uyc}, which demonstrated the computation of PDFs and the hadronic tensor using the $(1+1)$-dimensional Thirring model on a universal quantum computer. More recently, a simulation of the DIS hadronic tensor in the Schwinger model was performed using tensor networks~\cite{Ikeda:2025bjb}, enabling the extraction of the longitudinal structure function. In this work, rather than focusing on DIS structure functions, we investigate the low-$Q^2$ region of the hadronic tensor to extract hadron form factors, thereby providing a crucial complement to these developments. Furthermore, while the Schwinger-model DIS study of Ref.~\cite{Ikeda:2025bjb} employs staggered fermions, we emphasize the necessity of utilizing Wilson fermions. For the present form factor calculation, the Wilson formulation offers a distinct advantage because it preserves charge conjugation symmetry at finite lattice spacing. This allows the relevant hadronic states and current matrix elements to be classified more directly by their discrete quantum numbers, providing a cleaner setup for isolating elastic contributions and interpreting the form factor signals.

The objective of this work is therefore targeted yet concrete. We present a quantum-computing-oriented framework for evaluating the hadronic tensor from its real-time current-current correlation function in low-dimensional lattice gauge theories. This framework is implemented and benchmarked in both Abelian $\rm U(1)$ and non-Abelian $\rm SU(2)$ gauge theories in $(1+1)$ dimensions~\cite{Muschik:2016tws,Zache:2018jbt,Klco:2019evd,Calajo:2024qrc}. The $\rm U(1)$ theory serves as a controlled testbed via the Schwinger model, while the $\rm SU(2)$ theory admits gauge-singlet multi-fermion states that act as simple baryonic analogues~\cite{Atas:2021ext,Zhang:2024fgv,Lee:2024jnt}. Although these models do not reproduce the full spin, flavor, and spatial complexity of $(3+1)$-dimensional QCD, they share essential features with full gauge theories, including confinement mechanisms~\cite{Lowenstein:1971fc, Abdalla:1991vua, Coleman:1976uz} and chiral condensate formation~\cite{Byrnes:2002nv,Buyens2017fr,Zhang:2024fgv}. They therefore provide valuable benchmarks for developing and validating quantum algorithms aimed at hadron-structure observables in real world.

The formal definition of the hadronic tensor is common to both the low-$Q^2$ and DIS regimes, and the same real-time current-current correlator can in principle be evaluated at different momentum transfers. The restriction to low $Q^2$ in the present work is mainly practical, arising from the limited lattice size, momentum resolution, and real-time evolution that can be simulated efficiently on classical resources. Accessing the DIS regime would require larger spatial volumes, finer lattice spacings, higher momentum resolution, and more accurate long-time real-time evolution. These requirements are beyond the scope of the present classical simulations, but they are natural targets for future fault-tolerant quantum computers. In that setting, the same algorithmic framework may be extended toward high-$Q^2$ hadronic tensors and DIS structure functions. In the present work, we focus on the low-$Q^2$ response and demonstrate that the elastic form factors can be extracted from the computed hadronic tensor. The resulting form factors are compared with exact-diagonalization benchmarks, providing a direct test of the proposed real-time simulation strategy.

The rest of this paper is organized as follows. In Sec.~\ref{sec:LGT}, we introduce the Hamiltonian lattice formulations used in this work and describe the $\rm U(1)$ and $\rm SU(2)$ models with Wilson fermions. In Sec.~\ref{sec:quantum_algorithm}, we review the quantum algorithms for hadron-state preparation and for the simulation of current-current correlation functions. In Sec.~\ref{sec:result}, we present numerical results for the hadronic tensor and the extracted elastic form factors. Finally, we summarize our conclusions in Sec.~\ref{sec:summary}.

\section{the Lattice gauge theory in (1+1)-dimensions}\label{sec:LGT}
To simulate the hadronic tensor, the underlying gauge field theories must first be regularized on a lattice. For this purpose, we employ the Wilson fermion formalism~\cite{Wilson:1974sk}, which resolves the fermion doubling problem by introducing a Wilson term. This term explicitly lifts the unphysical doubler modes to the scale of the ultraviolet cutoff, giving them masses of order $\mathcal{O}(r/a_l)$, where $r$ is the Wilson parameter and $a_l$ denotes the lattice spacing~\cite{Zache:2018jbt}. In contrast to staggered fermions~\cite{Kogut1975hf}, the Wilson formulation cleanly preserves charge conjugation symmetry at finite lattice spacing. This property enables a rigorous classification of the target hadronic states by their $C$-parity~\cite{Chen:2025zeh}, which is crucial for interpreting the hadronic tensor where quantum selection rules associated with charge conjugation play a vital role.

In (1+1)-dimensions, the lattice Hamiltonian of Wilson fermions can be written in general as follows
\bea \label{eq: KS_ham}
    {H}_{W} =& -\frac{1}{2a_l}\sum^{N-1}_{n=0} \left[\bar{\psi}_n (r+i\gamma^1) \mathcal{U}_n {\psi}_{n+1} + { h.c.} \right] \nonumber\\
            &+   \left(m+\frac{r}{a_l}\right) \sum^{N-1}_{n=0} \bar{\psi}_n {\psi}_n + \frac{g^2 a_l}{2} \sum^{N-1}_{n=0} {\textbf{L}}^2_n\, ,
\eea
where $N$ is the number of lattice sites, $g$ is the bare coupling constant, $m$ is the bare mass, and ${\psi}_n$ is the two-component Dirac spinor at site $n$, $\psi_n = (\psi_{n,1},\psi_{n,2})^T$.
Throughout this work, the Wilson parameter is fixed to $r=1$, and we choose $\gamma^0=\sigma^1$, $\gamma^1=-i\sigma^3$ where $\sigma^1$ and $\sigma^3$ are Pauli matrices.
This convention leads to the Hamiltonian
\bea
    {H}_{W} =& -\frac{1}{a_l}\sum^{N-1}_{n=0} \left(\psi^\dagger_{n,2}\mathcal{U}_n \psi_{n+1,1} + { h.c.} \right) + \frac{g^2 a_l}{2} \sum^{N-1}_{n=0} {\textbf{L}}^2_n \nonumber\\
        &+ (m+\frac{1}{a_l}) \sum^{N-1}_{n=0}  \left( \psi^\dagger_{n,1} \psi_{n,2}+ h.c. \right) \, .
\eea
The gauge field degrees of freedom are encoded in the Wilson link operators $\mathcal{U}_n$, which ensure gauge invariance of the lattice Hamiltonian. 

Let $T^a$ be the generators of the Lie group $G$, which satisfy 
\bea
    [T^a,T^b]=if^{abc}T^c\,, 
\eea
where $f^{abc}$ are the structure constants of the group $G$. The conjugate momenta of $\mathcal{U}_n$ are denoted by the left or right electric fields $L^a_n$ and $R^a_n$, which are related by the adjoint representation, $R^a_n={(\mathcal{U}_n^{adj.})}_{ab} L^b_n$, with ${(\mathcal{U}_n^{adj.})}_{ab}= 2{\rm Tr}[\mathcal{U}_n T^a \mathcal{U}_n^{\dagger} T^b]$ and ${\textbf{L}_n^2}=\sum_a L^a_nL^a_n= \sum_a R^a_nR^a_n$. 
For a given lattice site $n$, operators $L^a$, $R^a$, and $\mathcal{U}$ satisfy the following commutation relations:
\begin{align}\label{eq:commu}
    [L^a,L^b] &= -if^{abc}L^c\,,   & \quad [R^a,R^b] &= if^{abc}R^c\,, \nonumber \\
    [L^a,\mathcal{U}] &= T^{a}\mathcal{U}\,,   & \quad [R^a,\mathcal{U}] &= \mathcal{U}T^a\,.
\end{align}
Eq.~\eqref{eq:commu} implies that the Hilbert space corresponding to the gauge link has infinite dimension. In (1+1)-dimensional gauge theories, the gauge fields can be eliminated by imposing Gauss's law~\cite{Hamer:1997dx, Sala:2018dui}. As a result, the gauge degrees of freedom can be integrated out, and the theory can be reformulated purely in terms of fermion fields. The lattice version of Gauss's law $G_n^a = 0$ is given by~\cite{Hamer:1997dx}
\bea\label{eq: Gauss1}
    {G^a_n}&\equiv {L^a_n}-{R^a_{n-1}}-{\mathcal{Q}^a_n} = 0\,,
\eea
where ${\mathcal{Q}}_n^a$ is the charge density on lattice site $n$, which only depends on the fermion field $\psi^\dagger_n$ and $\psi_n$.
Solving Eq.~\eqref{eq: Gauss1} requires a choice of boundary conditions. We adopt periodic boundary conditions (PBCs) to preserve translational symmetry, thereby ensuring that hadron momenta are well-defined. Under PBCs, the general solution of Gauss's law is given by 
\begin{gather}
    {L}^a_n = \mathcal{E}^a+\delta L^a_n =\mathcal{E}^a + \frac{1}{N}\sum^{N}_{n'=1} \left(n'-N\theta_{n'>n}\right)\mathcal{Q}^a_{n'}\,, \nn
    \theta_{n'>n}=\left\{
    \begin{aligned}\label{eq:eletric}
        1 \,&,\, n'>n\,,\\
        0 \,&,\, {\rm others}\,,
    \end{aligned}
    \right.
\end{gather}
where $\mathcal{E}^a=\frac{1}{N}\sum_n L^a_n$ is the average electric fields and it is an undetermined constant. In this work, we select $\mathcal{E}^a=0$ as a boundary condition to determined that constant. After that, $L_n^a$ can be represented by $\delta L_n^a$, which is the fluctuation around average color electric field. After solving $\delta L_n^a$, the $H_{W}$ depends only on the fermion fields $\psi_n$. 

The above discussions are general, and the charge operator $\mathcal{Q}^a_n$ for specific cases such as $\rm U(1)$ and $\rm SU(2)$ gauge field theory can be written explicitly. In the Schwinger model, the only generator of the $\rm U(1)$ group is $T^1=1$, so the Abelian charge is defined as ${\mathcal{Q}}_n=\sum_s{\psi}^\dagger_{n,s}{\psi}_{n,s}-I$. The total charge $\mathcal{Q}_{tot}=\sum_{n=0}^{N-1}\mathcal{Q}_n$ is a conserved quantity. Combining with translation invariance of PBCs, we can use three quantum numbers to label the energy eigenstate of the Schwinger model, i.e. energy $E$, momentum $\mathbf{p}$, and electric charge $\mathcal{Q}_{tot}$.

In the non-Abelian $\rm SU(2)$ case, each spinor of the fermion field consists of red and green components labeled by $c=0,1$, $\psi_{n,s} = (\psi^0_{n,s},\psi^1_{n,s})^T$, so it spans the fundamental representation space of the $\rm SU(2)$ gauge group. In this representation, the three generators are $T^a=\frac{1}{2}\sigma^a$, where $\sigma^a$ represents the three Pauli matrices, and the structure constants are given by the Levi-Civita tensor $f^{abc}=\epsilon^{abc}$. The non-Abelian charges are defined as $\mathcal{Q}^a_n= \sum_s\psi_{n,s}^{\dagger}T^a\psi_{n,s}$, and the total charge $\mathcal{Q}^a_{tot}=\sum_{n=0}^{N-1}\mathcal{Q}^a_n$ are conserved quantities. 
In addition to the three non-Abelian charges, the system possesses another conserved quantity, baryon number $B$, arising from global $\rm U(1)$ symmetry, which is defined as
\bea
    B=-N+\frac{1}{2} \sum_{n=0}^{N-1}\sum_s\left({\psi_{n,s}^{0\dagger}}\psi_{n,s}^0+{\psi_{n,s}^{1\dagger}}\psi_{n,s}^1 \right)\,.
\eea
Therefore, the good quantum numbers in $\rm SU(2)$ LGT are $E$, $\mathbf{p}$, $(\mathcal{Q}_{tot})^2 \equiv \sum_a (\mathcal{Q}^a_{tot})^2$, $\mathcal{Q}^3_{tot}$, and $B$. All of the energy eigenstates in $\rm SU(2)$ LGT can be labeled by those five quantum numbers.

To simulate the system on a quantum computer, the Hamiltonian must be expressed in terms of Pauli operators. This can be achieved by the Jordan-Wigner transformation~\cite{Backens_2019}
\bea\label{eq:JwT}
    {\phi}_k \rightarrow \left(\prod_{n<k} {\sigma}^3_{ n}\right){\sigma}^-_{ k} \,,\quad
    {\phi}_k^{\dagger} \rightarrow \left(\prod_{n<k} {\sigma}^3_{ n}\right){\sigma}^+_{k}\,,
\eea
where $\sigma^i_{ n}$ represents the Pauli matrices acting on the n-th qubit, and ${\sigma}^{\pm}=({\sigma}^1\pm i{\sigma}^2)/2$ . This transformation preserves the canonical anticommutation relations of the fermion field.

For the $\rm U(1)$ case, the fermion field $\phi_k=\psi_{n,s}$ carries a site index $n$, and a spinor index $s=1,2$, is mapped to a qubit with index $k=2n+s-1$. Following this and the Jordan-Wigner transformation, the Abelian charge ${\mathcal{Q}}_n$ is given by 
\bea
    {\mathcal{Q}}_n&= \frac{1}{2}\left(\sigma^3_{2n}+\sigma^3_{2n+1}\right)\,.
\eea
The kinematic, mass, and electric energy terms in the Hamiltonian can be written as
\bea
    {H}_{\rm{kin}} &=  \sum^{N-1}_{n=0}\left( {\sigma}^+_{2n+1}{\sigma}^-_{2n+2}+h.c.\right)\,,\nn
    {H}_{\rm{m}} &= -(m+1)  \sum^{N-1}_{n=0}\left( {\sigma}^+_{2n}{\sigma}^-_{2n+1}+h.c.\right)\,,\nn
    {H}_{\rm{e}}&=\frac{g^2 a_l}{2N^2} \sum^{N-1}_{n=0} \left[ \sum^{N}_{n'=1} (n'-N\theta_{n'>n})\mathcal{Q}_{n'} \right]^2.
\eea

For the $\rm SU(2)$ LGT, the fermion field $\phi_k=\psi^c_{n,s}$ carries an additional color index $c$, and is mapped to a qubit labeled by $k=4n+2(s-1)+c$. After that, the non-Abelian charges are given by
\bea\label{eq:pauQ}
    \mathcal{Q}^1_n  =& -\frac{1}{2}\left[ (\sigma^+_{4n}\sigma^-_{4n+1}+\sigma^+_{4n+2}\sigma^-_{4n+3}) +h.c. \right]\,,\nn
    \mathcal{Q}^2_n  =& \frac{i}{2}\left[ (\sigma^+_{4n}\sigma^-_{4n+1}+\sigma^+_{4n+2}\sigma^-_{4n+3}) - h.c. \right]\,,\nn
    \mathcal{Q}^3_n  =& \frac{1}{4}\left(\sigma^3_{4n}+\sigma^3_{4n+2}-\sigma^3_{4n+1}-\sigma^3_{4n+3}\right)\,.
\eea 
The qubit Hamiltonian of $\rm SU(2)$ LGT can be obtained by combining the Eq.~\eqref{eq: KS_ham}, Eq.~\eqref{eq:eletric}, Eq.~\eqref{eq:JwT} and Eq.~\eqref{eq:pauQ}. Then the Hamiltonian is given by $\bar{H}=\bar{H}_{\rm{kin}}+\bar{H}_{\rm{m}}+\bar{H}_{\rm{e}}$, with
\bea
    \bar{H}_{\rm{kin}}= \,& \sum^{N-1}_{n=0}\left[({\sigma}^+_{4n+2}{\sigma}^3_{4n+3}{\sigma}^-_{4n+4} \right. \nn & \left. + {\sigma}^+_{4n+3}{\sigma}^3_{4n+4}{\sigma}^-_{4n+5})+h.c.\right]\,,\nn
    \bar{H}_{\rm{m}}=\,&-(m+1)\sum^{N-1}_{n=0}\left[({\sigma}^+_{4n}{\sigma}^3_{4n+1}{\sigma}^-_{4n+2} \right. \nn & \left. + {\sigma}^+_{4n+1}{\sigma}^3_{4n+2}{\sigma}^-_{4n+3})+h.c.\right]\,,\nn
    \bar{H}_{\rm{e}}=\, & \frac{g^2 a_l}{2N^2}\sum_a \sum^{N-1}_{n=0} \left[ \sum^{N}_{n'=1} (n'-N\theta_{n'>n})\mathcal{Q}_{n'}^a \right]^2\,.
\eea

The hadronic tensor in (1+1)-dimensions also needs to be mapped to qubits. This can be done by the Jordan-Wigner transformation:
\bea\label{eq:HT_dis}
    W^{\mu \nu}(q^0,q^1)= \int dt \sum_{z} e^{iq^0t}\,e^{-iq^1z}\,\mathcal{O}^{\mu\nu}(z,t)\,,
\eea
where
\bea
    \mathcal{O}^{\mu\nu}(z,t)=\bra{h}  e^{iHt} J^{\mu}_m \,e^{-iHt} J^\nu_n  \ket{h}\,,\quad z=m-n\,.
\eea
The current operator $J^\mu_n$ can then be expressed in terms of Pauli operators as follows.
For the $\rm U(1)$ LGT:
\bea
    {J^0_n}&=\frac{e_f}{2}\left(\sigma^3_{2n}+\sigma^3_{2n+1}\right)\,,\nn
    {J^1_n}&=ie_f\left({\sigma}^+_{{2n+1}}{\sigma}^-_{{2n}}-{\sigma}^+_{{2n}}{\sigma}^-_{{2n+1}}\right)\,,
\eea
 and for $\rm SU(2)$ LGT:
 \bea
    {J^0_n}&=\frac{e_f}{2}\left(\sigma^3_{4n}+\sigma^3_{4n+1}+\sigma^3_{4n+2}+\sigma^3_{4n+3}\right)\,, \nn
    {J^1_n}&=ie_f\left[(\sigma^+_{4n+2}\sigma^3_{4n+1}\sigma^-_{4n}+\sigma^+_{4n+3}\sigma^3_{4n+2}\sigma^-_{4n+1})-h.c.\right]\,.
\eea
With the above mapping of lattice gauge theory to qubits, we are ready to simulate the hadronic tensor on a quantum computer.

\section{Quantum algorithms}\label{sec:quantum_algorithm}
By employing the quantum algorithm proposed in Ref.~\cite{Li:2021kcs}, the simulation of the hadronic tensor on a quantum computer proceeds in two main steps: 1. Preparation of the interacting vacuum state $\ket{\Omega}$ and the hadron state $\ket{h}$; 2. Evaluation of the current-current correlation functions. 

The hadron state can be prepared by a quantum-number-resolving variational quantum eigensolver (VQE). The framework of this method has been discussed in Ref.~\cite{Li:2023kex}, and we give a brief review here. 
Consider the target hadron state as the $k$-th excited state with quantum numbers $l$.
To prepare the lowest $k$ states simultaneously.
We need $k$ initial reference states $\ket{\psi_{l,i}}_{\rm ref}$, $i=0,1,...,k-1$, which have the same quantum number $l$ as the hadron state. Then, a quantum number preserving ansatz $\rm{U}_A(\theta)$ is needed to generate the trial wave functions $\ket{\psi_{l,i}(\theta)}\equiv {\rm{U}}_A(\theta)\ket{\psi_{l,i}}_{\rm ref}$, which have the same quantum number as the hadron state. $\rm{U}_A(\theta)$ can be constructed by quantum alternating operator ansatz (QAOA). The key point of the QAOA lies in splitting the Hamiltonian into $d$ ($d\geq2$) parts, $H=H_1+\cdots+H_d$, while satisfying $[H_i, H_{i+1}]\not=0$ and $[H_i, \mathcal{C}]=0$ for every $H_i$ and conserved charge $\mathcal{C}$. Then $\rm{U}_A(\theta)$ is given by
\bea\label{eq:U}
    {\rm{U}_A}(\theta) = \prod_{i=1}^p \left(\prod_{j=1}^d  e^{i\theta_{ij}H_j}\right)\,,
\eea
where $p$ is the number of layers for alternating evolution of the Hamiltonian.

It is necessary to minimize the cost function $E_l(\theta)$ on the classical computer to select the target hadron state from numerous trial states~\cite{Nakanishi2019ss},
\bea\label{eq:costfunc}
    E_l(\theta) = \sum_{i=0}^{k-1} w_{li}\bra{\psi_{l,i}(\theta)}H\ket{\psi_{l,i}(\theta)}\,,
\eea
where the weights $w_{li}$ are required to satisfy $w_{l,0}>w_{l,1}>\dots>w_{l,k-1}>0$, so that 
${\rm{U}_A}(\theta^*)\ket{\psi_{l,i}}_{\rm ref}$ are the eigenstates in quantum numbers $l$ with the optimized parameter $\theta^*$. The $(k-1)$-th excited hadron state $\ket{h}$ is then prepared as $\ket{h}={\rm{U}_A}(\theta^*)\ket{\psi_{l,k-1}}$.

In the case of the $\rm U(1)$, the conserved quantities include charge and momentum, so the $\rm{U}_A(\theta)$ needs to preserve those two quantum numbers. The vacuum and the lightest $q\bar{q}$ bound state are the ground and first excited state in the $\mathcal{Q}_{tot}=0$ and $\mathbf{p}=0$ subspace. 
The construction of the reference states and the decomposition of the Hamiltonian for $\rm U(1)$ follow a procedure analogous to that of the $\rm SU(2)$ case discussed below. 

As discussed in Sec.~\ref{sec:LGT}, the quantum number of $\rm SU(2)$ LGT is more complex than $\rm U(1)$ LGT. That means we need to design the reference states and divide the Hamiltonian more carefully. The vacuum and the lightest $q\bar{q}$ meson state considered here are the ground and first excited states in the $(\mathcal{Q}_{tot})^2=\mathcal{Q}^3_{tot}=0$, $B=0$, and $\mathbf{p}=0$ subspace. We are also concerned about the lightest $qq$ baryon bound state, which is the ground state of the $(\mathcal{Q}_{tot})^2=\mathcal{Q}^3_{tot}=0$, $B=1$, and $\mathbf{p}=0$ subspace. Except for baryon number $B$ and energy $E$, the remaining quantum numbers of the above states are the same. To construct suitable reference states within these sectors, we first identify local color-singlet subspaces and then select states with fixed baryon number.
We use $B$ and $i$ to label the local reference states $\ket{\psi_{B,i}}_{\rm local \,\,ref}$:
\bea\label{eq:reference state}
   \ket{\psi_{0,0}}_{\rm local \,\,ref} =& \frac{1}{2}\left( \ket{ 0011} - \ket{0110} +\ket{1001} + \ket{1100}\right)\,,\nn
   \ket{\psi_{0,1}}_{\rm local \,\,ref} =&  \frac{1}{\sqrt{2}}\left( \ket{ 0011} - \ket{1100}\right)\,,\nn
   \ket{\psi_{1,0}}_{\rm local \,\,ref} =&  \ket{ 0000} \,,
\eea
where the states on the right-hand side of the equation are the computational basis, which are defined by:
\bea
    \ket{i_0,i_1,\cdots,i_{N-1}}&\equiv \ket{i_0}\otimes\ket{i_1}\otimes\cdots\otimes\ket{i_{N-1}}\,,\nn
    i_0,\cdots,i_{N-1}&=0,1\,,
\eea
where $\sigma^3\ket{0}\equiv\ket{0}$, $\sigma^3\ket{1}\equiv-\ket{1}$.
Therefore, the reference states are constructed as tensor products of these local reference states:
\bea
    \ket{\psi_{B,i}}_{\rm ref} 
    = & \frac{1}{\sqrt{N}} \sum_{n=1}^{N}  \ket{\psi_{0,0}}_{\rm local \,\,ref}^{\otimes (n-1)}   \nn 
    & \otimes \ket{\psi_{B,i}}_{\rm local \,\,ref} \otimes \ket{\psi_{0,0}}_{\rm local \,\,ref}^{\otimes (N-n)} \,.
\eea
It is straightforward to verify that those three reference states have the same quantum number (except energy $E$) of the target vacuum, meson, and baryon state. 

\begin{figure}[htbp]
  \centering
  
  \subfigure[~Detailed quantum circuit for operator $\rm U_{0,0}$.]{
    \includegraphics[width=0.4\textwidth]{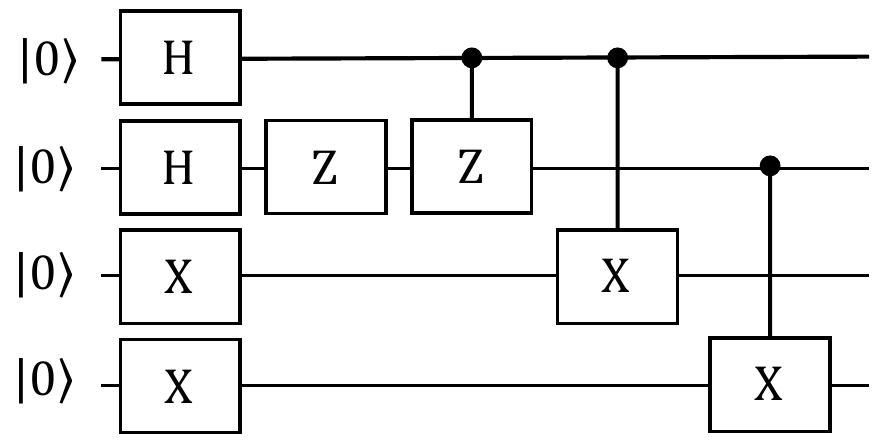}
    \label{fig:ref state vac}
  }
  \subfigure[~Detailed quantum circuit for operator $\rm U_{0,1}$.]{
    \includegraphics[width=0.4\textwidth]{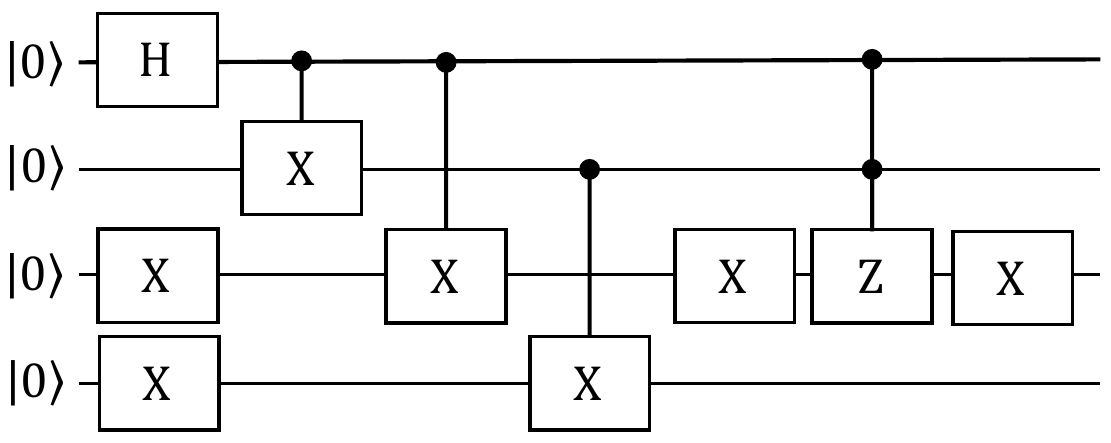}
    \label{fig:ref state meson}
  }
  \subfigure[~Quantum circuit for preparing reference states $\ket{\psi_{0,1}}_{\rm ref}$ from the Dicke state $\ket{D^{N}_{1}}$.]{
    \includegraphics[width=0.45\textwidth]{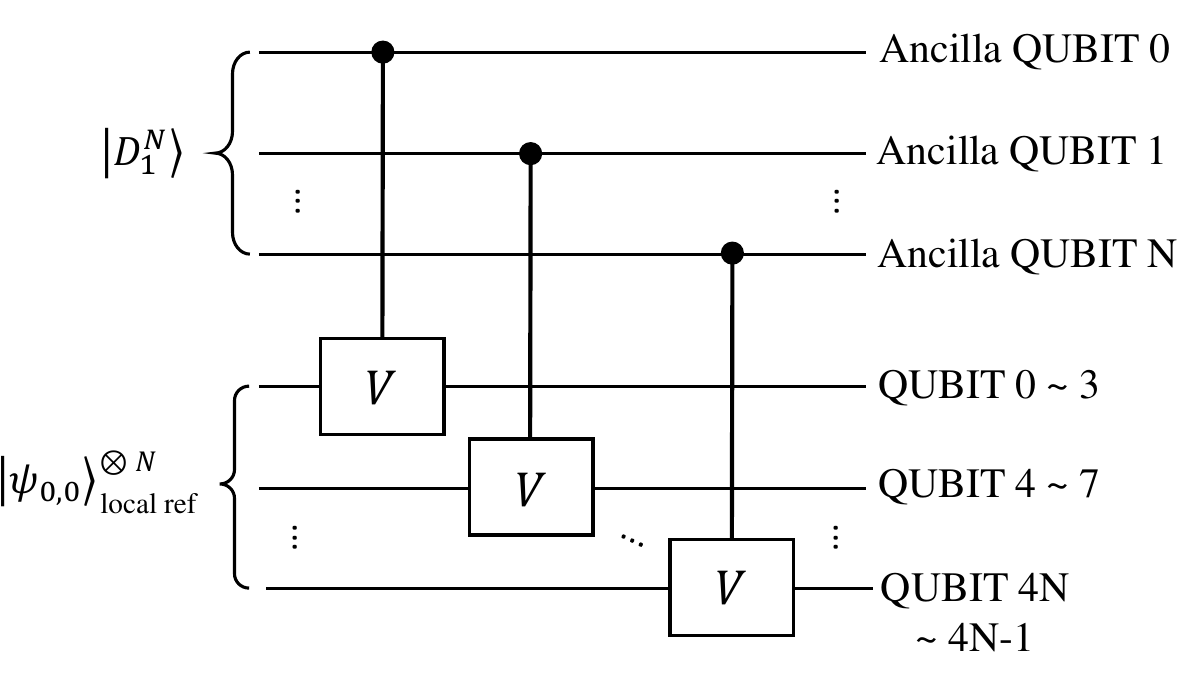}
    \label{fig:reference state}
  }
    \caption{Quantum circuits for preparing the reference states. }
  \label{fig: quantum circuit reference state}
\end{figure}

To implement this state, we adopt a three-step procedure.
First, we initialize a set of $N$ ancilla qubits in the Dicke state $\ket{D^{N}_1}$~\cite{Bartschi:2019zce}. 
In general, the Dicke state $\ket{D^n_l}$ is defined as the equal superposition of all $n$-qubit states $\ket{x}$
\bea
    \ket{D^n_l}\equiv \sqrt{\frac{1}{C^l_n}}\sum_{\substack{x\in \{0,1\}^n \\ w(x)=l}}\ket{x}\,,
\eea
where $C^l_n=\frac{n!}{l!(n-l)!}$ and $w(x)$ is the Hamming weight, representing the number of ones in $\ket{x}$. 
Second, we define two unitary operators, $\rm U_{0,0}$ and $\rm U_{0,1}$, acting on four qubits, such that $\rm U_{0,0}\ket{0}^{\otimes4}=\ket{\psi_{0,0}}_{\rm local \,\,ref}$ and $\rm U_{0,1} \ket{0}^{\otimes4}=\ket{\psi_{0,1}}_{\rm local \,\,ref}$. 
The detailed quantum circuits for $\rm U_{0,0}$ and $\rm U_{0,1}$ are shown in Fig.~\ref{fig:ref state vac} and Fig.~\ref{fig:ref state meson}.
A transition operator $\rm V=\rm U_{0,1}\rm U_{0,0}^{\dagger}$ is then constructed to map the state $\ket{\psi_{0,0}}_{\rm local \,\,ref}$ to $\ket{\psi_{0,1}}_{\rm local \,\,ref}$. 
The $\rm U_{0,0}^{\otimes N}$ operator is then applied to prepare the product state $\ket{\psi_{0,0}}_{\rm local \,\,ref}^{\otimes N}$.
Finally, as shown in Fig.~\ref{fig:reference state}, $\ket{\psi_{0,1}}_{\rm ref}$ can be prepared by applying a controlled-$\rm V$ gate from the $i$-th ancilla qubit in $\ket{D^{N}_{1}}$ to the corresponding $i$-th local reference state $\ket{\psi_{0,0}}_{\rm local\,\, ref}$. 
The state $\ket{\psi_{1,0}}_{\rm ref}$ can be prepared similarly by replacing all controlled-${\rm V}$ gates with controlled-${\rm U}_{0,0}^{\dagger}$ gates.

In addition, the Hamiltonian of $\rm SU(2)$ can be divided as follows: 
\bea\label{eq:dividedham} 
    {H}_1= \frac{1}{2}\bar{H}_{\rm{kin}}\,,\,
    {H}_2=\bar{H}_{\rm{m}}\,,\,
    {H}_3= \frac{1}{2} \bar{H}_{\rm{kin}}\,,\,
    {H}_4=\bar{H}_{\rm{e}}\,.
\eea
Note that each term in Eq.~(\ref{eq:dividedham}) satisfies 
\bea
    \left[H_i,H_{i+1}\right]\not = 0\,,\quad\left[\mathcal{C},H_i\right] = 0\,,\quad \left[T,H_i\right] = 0\,,
\eea
where $\mathcal{C}=(\mathcal{Q}_{tot})^2,\,\mathcal{Q}^3_{tot},\,B$, and $T$ is the translation operator. Thus, the decomposition of the Hamiltonian is feasible. Then we set $p=N$ in Eq.~(\ref{eq:U}) and $w_{B,0}=1$, $w_{B,1}=0.5$ in Eq.~(\ref{eq:costfunc}) to prepare the state in $B=0$ sector. 
After optimizing the cost function, the vacuum state and the hadron state can be prepared by ${\rm{U}_A}(\theta^*_{B})$
\bea
    &\ket{\Omega} = {\rm{U}_A}(\theta^*_{0})\ket{\psi_{0,0}}_{\rm{ref}}\,,\nn
    &\ket{h=q\bar{q}} = {\rm{U}_A}(\theta^*_{0})\ket{\psi_{0,1}}_{\rm{ref}}\,,\nn
    &\ket{h=qq} = {\rm{U}_A}(\theta^*_{1})\ket{\psi_{1,0}}_{\rm{ref}}\,.
\eea

\begin{figure}[htbp]
  \centering
  
  \subfigure[~Real part of correlation $\bra{h}\mathcal{O}\ket{h}$.]{
    \includegraphics[width=0.4\textwidth]{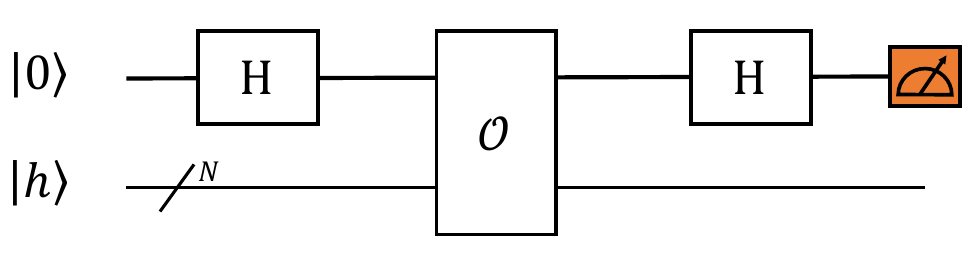}
    \label{fig:circuit sub1}
  }
  \subfigure[~Detailed quantum circuit for the implementation of operator $\mathcal{O}$.]{
    \includegraphics[width=0.45\textwidth]{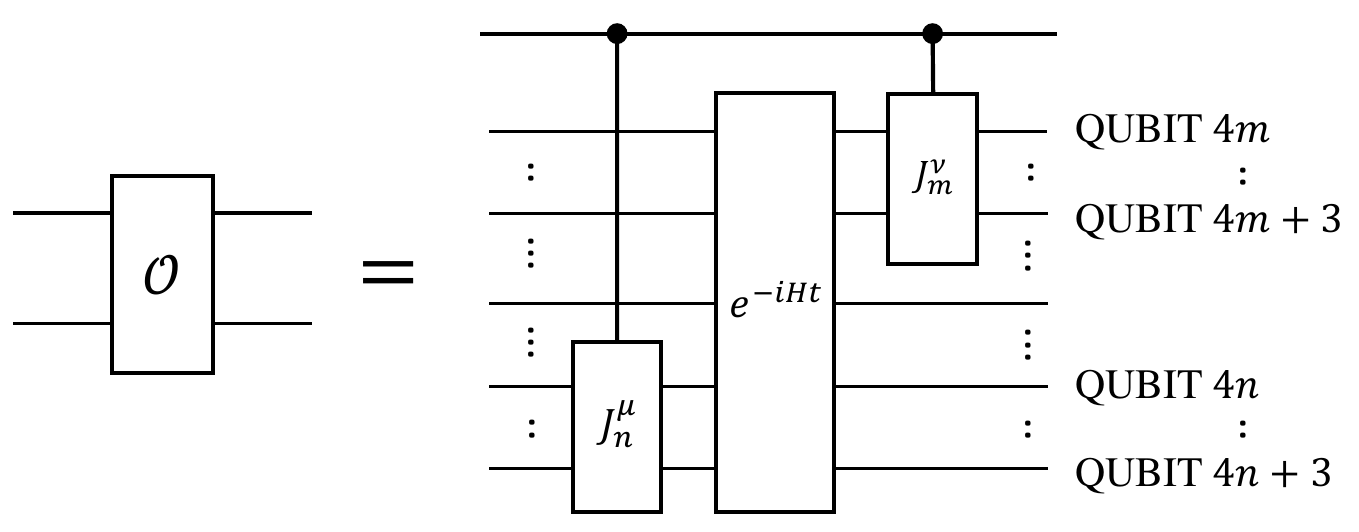}
    \label{fig:circuit sub2}
  }
  \caption{Quantum algorithm for computing current-current correlation functions. }
  \label{fig:quantum circuit}
\end{figure}

After preparing the hadron state, we perform the quantum algorithm, which is shown in Fig.~\ref{fig:circuit sub1}, to simulate the real part of current-current correlation functions by using an ancilla qubit~\cite{Pedernales:2014izf}.
The imaginary part can be obtained by replacing the last Hadamard gate with the $R_x(\pi/2)$ gate on the ancilla qubit in Fig.~\ref{fig:circuit sub1}. 
Furthermore, the time evolution operator in Fig.~\ref{fig:circuit sub2} can be implemented on a quantum computer using the standard Trotterization method~\cite{nielsen_chuang_2010}. 

Finally, to obtain the hadronic tensor $W^{\mu\nu}(q)$, one just need to perform a Fourier transformation on the simulated current-current correlation function, which can be done numerically on a classical computer.

\section{Results}\label{sec:result}

We use open-source package QuSpin~\cite{Phillip2017qa} for the quantum simulation on classical computers. In particular, we simulate the hadronic tensor of the lowest-lying meson in the $\rm U(1)$ LGT using $N_q=22$ qubits, and both meson and baryon states in the $\rm SU(2)$ LGT using $N_q=20$ qubits. 

First, we select the bare parameters $m$ and $g$ to prepare the zero-momentum hadron state $\ket{h}$ with mass defined as $m_h=\bra{h}H\ket{h}-\bra{\Omega}H\ket{\Omega}$. The dimensionless mass $m_h a_l$ must be chosen to keep the lattice spacing and finite-volume effects under control. 
In practice, the simulation parameters and resulting hadron masses are summarized in Table~\ref{table_para}.
In the SU(2) case, the meson and baryon masses are found to be degenerate due to the charge conjugation symmetry of the theory.
\begin{table} 
  \centering
  \caption{Choice of bare parameters and resulting hadron masses.}
  \begin{tabularx}{\linewidth}{*{6}{>{\centering\arraybackslash}X}}
    \toprule
     Theory & $m$ & $g$ & $e_f$ & $m_{\rm M }a_l$ & $m_{\rm B}a_l$ \\
    \midrule
    U(1) & 0.25 & 1 & 1 & 1.04 & -- \\
    SU(2) & 0.3 & 1 & 1 & 0.83 & 0.83 \\
    \bottomrule
  \end{tabularx}
  \label{table_para}
\end{table}

\subsection{Hadronic tensor of $q\bar{q}$ meson state in $\rm U(1)$ LGT}
\begin{figure}[htbp]
	\centering
	\includegraphics[width=0.45\textwidth]{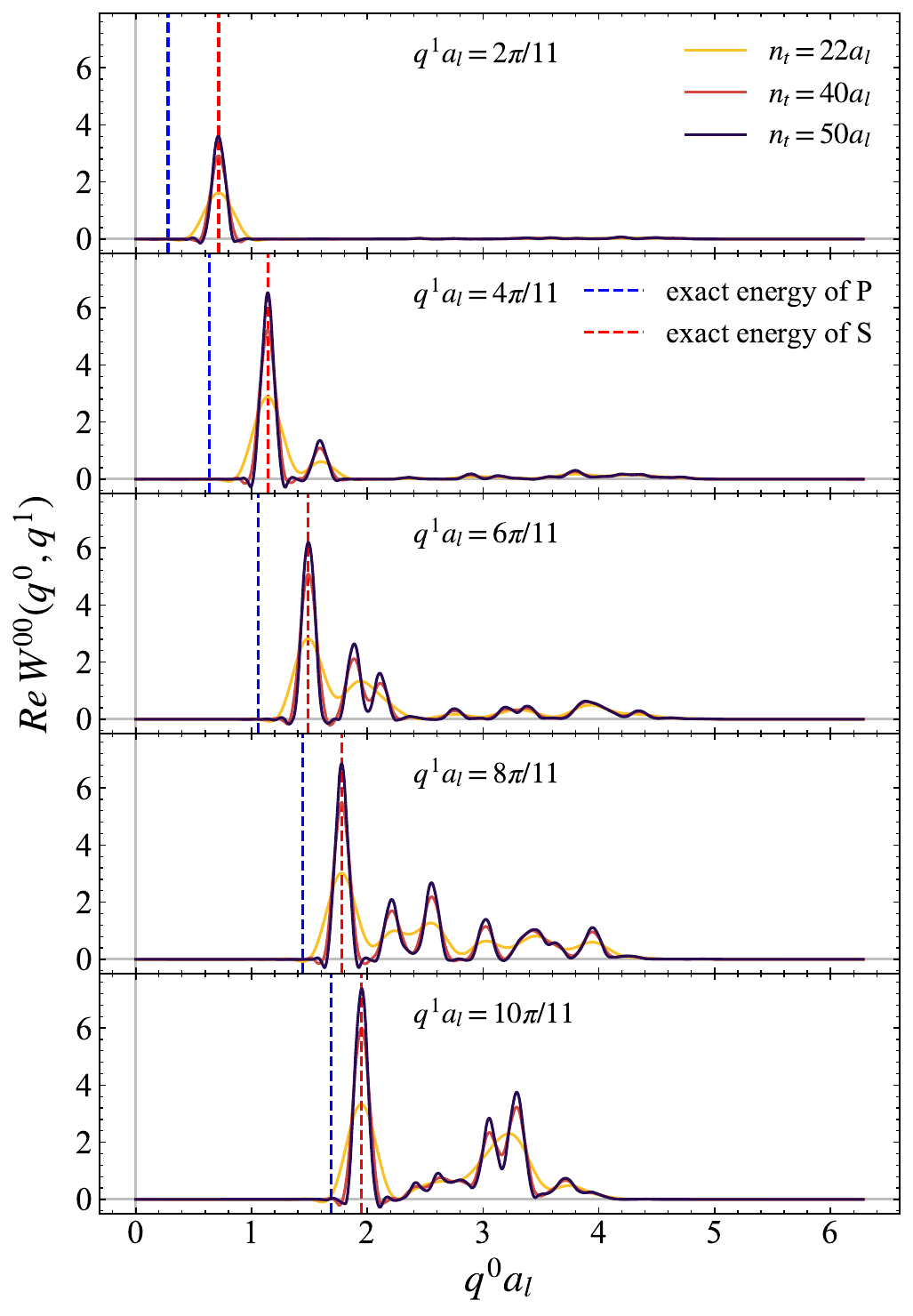}
	\caption{$W^{00}(q)$ of $\rm U(1)$ LGT as a function of $q^0a_l$ and $q^1a_l$, with $N_q=22$, $m=0.25$, $g=1$, and $m_{\rm M} a_l=1.04$. Dashed lines denote the exact energy $\Delta E= p^{\prime 0}-p^0_M$ of the lowest-lying pseudoscalar (P) and the scalar (S) meson states obtained from the Hamiltonian spectrum.
	}
	\label{fig: W00}
\end{figure}

After fixing the parameters $m$, $g$, and $m_{\rm M}a_l$ in the $\rm U(1)$ LGT, we can simulate the quantum algorithm on a classical device. We show the real part of the hadronic tensor $W^{00}(q)$ for the lightest $q\bar{q}$ bound state in $\rm U(1)$ LGT in Fig.~\ref{fig: W00}, as a function of continuous $q^0a_l$ and discrete $q^1a_l$. The imaginary part of $W^{00}$ is nearly zero and not shown in the figure. In the Hamiltonian formulation of LGT, space is discretized while time remains continuous. Consequently, the spatial momentum $q^1a_l$ takes discrete values $q^1a_l = 2\pi k/N,$ where $k=0,1,...,N-1$. 
For a given $q^1a_l$, the variable $q^0a_l$ remains continuous, as it is conjugate to the continuous time $t$.

From Fig.~\ref{fig: W00}, we can see that for a given $q^1a_l$, the hadronic tensor $W^{00}(q)$ has a series of peaks in different values of $q^0 a_l$. 
This structure arises from the spectral decomposition. By inserting the completeness relation of the Hilbert space into the current-current correlation function, we obtain:
\bea\label{eq:ht definition}
    W^{\mu\nu}(q)=&\sum_X\int d\Pi_X(2\pi)^2\delta^{(2)}(q+p_h-p_X) \,\nn
    &\times\bra{h}J^{\mu}(0)\ket{X}\bra{X}J^{\nu}(0)\ket{h}\,,
\eea
where $d\Pi_X$ is the phase space of the final states $\ket{X}$ and the spacetime translation of the matrix element $\bra{h}J^{\mu}(z)\ket{X}=\bra{h}J^{\mu}(0)\ket{X}e^{i(p_h-p_X)z}$ was used. 
Eq.~(\ref{eq:ht definition}) implies $W^{\mu\nu}(q)$ should have a nonzero contribution when the delta function is satisfied for $q=p_X-p_h$. 
After choosing the current component $\mu=\nu=0$ and isolating the contribution from the one-particle final states $\ket{h'(p')}$, we have
\bea\label{eq:ht_onep}
    W&^{00}(q)=\,\sum_{h'}\frac{2\pi}{2E_{p^\prime}}\delta^{(1)}(q^0+p^0-p^{\prime 0})\big|_{{q^1}={p^{\prime1}}-{p^1}} \nn
    &\times\bra{h}J^{0}(0)\ket{h'(p')}\bra{h'(p')}J^{0}(0)\ket{h} + \text{others}\,,
\eea
where ``others" denotes the contributions from multiple-particle final states.
As shown in Eq.~(\ref{eq:ht_onep}), the terms with final states being one-particle states contribute a sum of delta functions to the hadronic tensor, which appear as sharp peaks whenever the kinematic variables satisfy energy-momentum conservation.
In Fig.~\ref{fig: W00}, dashed lines denote the exact energy $\Delta E= p^{\prime 0}-p^0_M$ of the lowest-lying pseudoscalar (P) and the scalar (S) meson state of the lattice Schwinger model. Notice that the electromagnetic current is odd under charge conjugation, $  C J^\mu C^{-1} = - J^\mu $, and the lightest $q\bar{q}$ bound state is even under charge conjugation. This implies that only C-even intermediate states can contribute to the hadronic tensor.
Consequently, the C-odd pseudoscalar meson does not contribute at the corresponding pseudoscalar meson energy, whereas the C-even scalar meson gives a nonzero contribution. This selection rule relies on charge conjugation symmetry, which is preserved in our Wilson-fermion formulation. While we have checked numerically that this symmetry is violated in Staggerd fermion formalism.
We emphasize that, although the momentum transfer accessible in the present calculations is limited and remains of order $Q^2 \sim m_h^2 $, all contributions shown in Fig.~\ref{fig: W00} are inelastic, since the final states $\ket{X}$ differ from the initial hadron $\ket{h}$. Therefore, no elastic contribution ($X=h$) appears in the hadronic tensor in the present calculation. 
Moreover, one can see that the first peak of $W^{00}(q^0,q^1)$ moves to larger values of $q^0a_l$ as $q^1a_l$ increases; this is because the $p^{\prime 0}a_l$ of one-particle state increases when it has larger and larger spatial momentum.
At higher energies, contributions from multi-particle final states become significant.

In Eq.~\eqref{eq:HT_dis}, the integration over time $t$ runs from $-\infty$ to $\infty$. For numerical evaluation, a finite cutoff $(-n_t,\,n_t\,)$ in $t$ is required, which replaces the exact delta function with a broadened peak. We multiply by a Hann window function in the time domain to reduce boundary effects~\cite{Harris:1978yaq,White:2004etd}. As shown in Fig.~\ref{fig: W00}, the peak width decreases while its height increases as $n_t$ grows.

\subsection{Hadronic tensor of $q\bar{q}$ meson and $qq$ baryon states in $\rm SU(2)$ LGT}

\begin{figure}[htbp]
  \centering
  
  \subfigure[~$W^{00}(q)$ of meson state in SU(2) LGT.]{
    \includegraphics[width=0.45\textwidth]{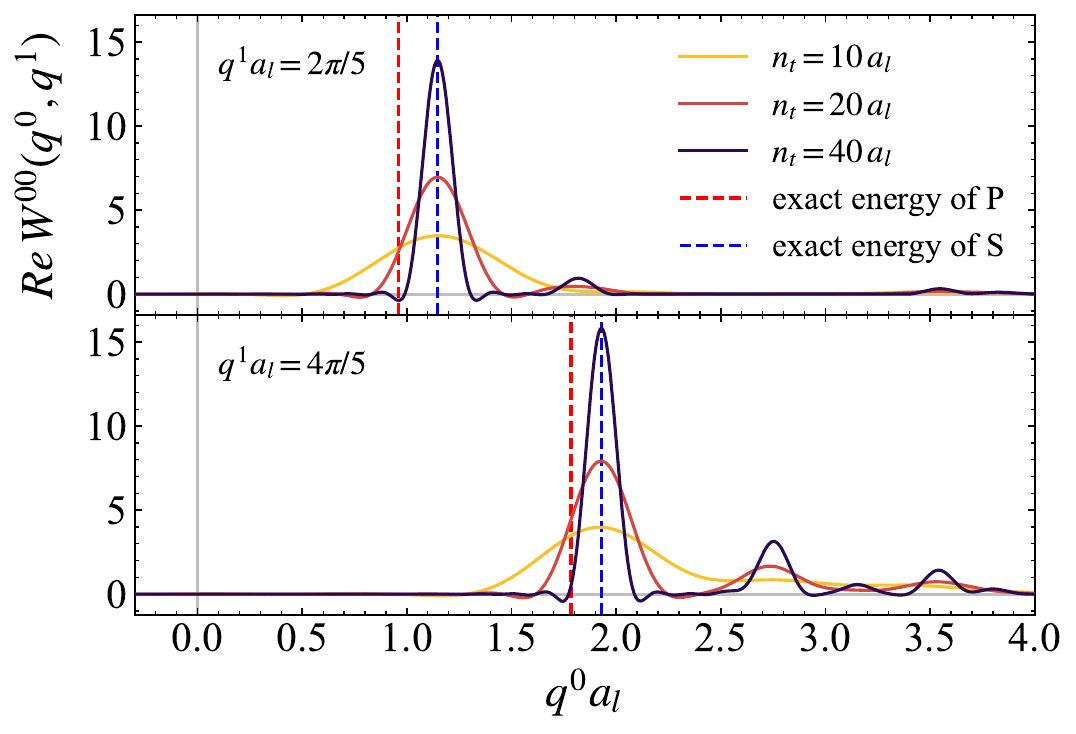}
    \label{fig:meson_ht}
  }
  \subfigure[~$W^{00}(q)$ of baryon state in SU(2) LGT.]{
    \includegraphics[width=0.45\textwidth]{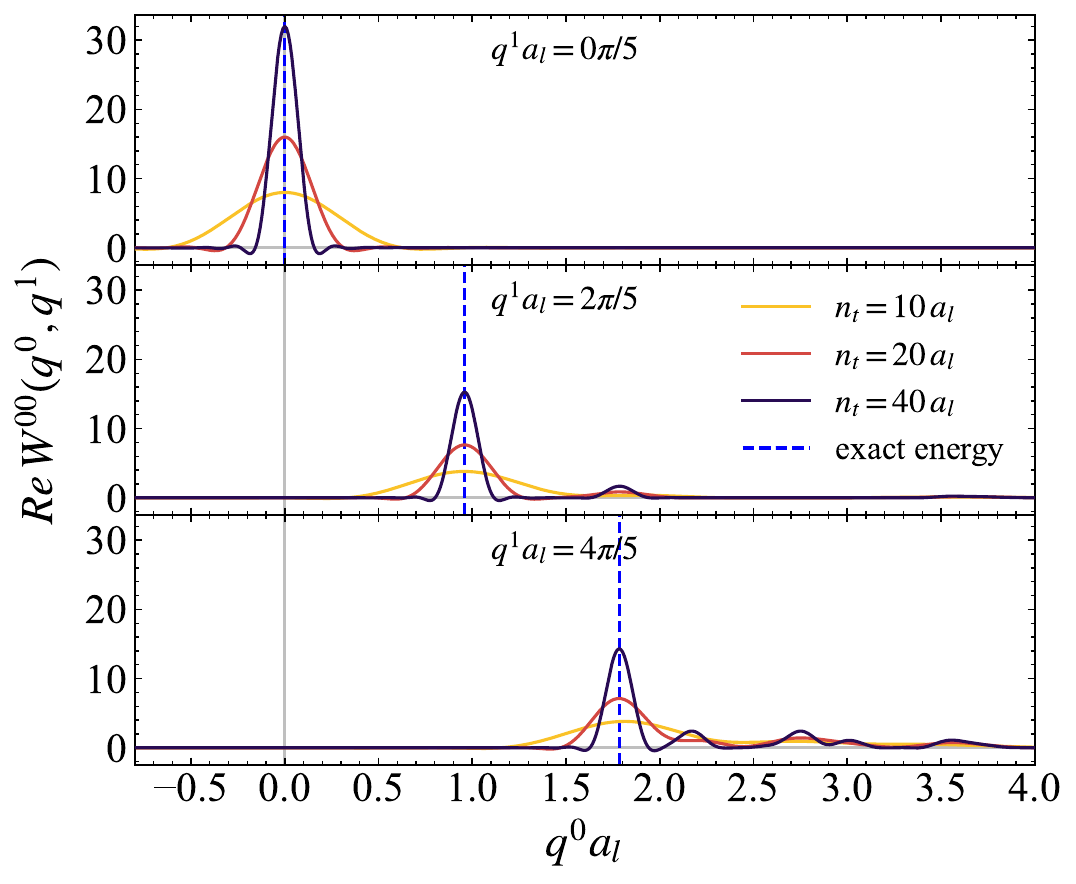}
    \label{fig:baryon_ht}
  }
  \caption{$W^{00}(q)$ of $\rm SU(2)$ LGT as functions of $q^0$ and $q^1$, with $N_q=20$, $g=1$, and $m=0.30$, meson mass $m_{\rm{M}}a_l=0.83$ and baryon mass $m_{\rm{B}}a_l=0.83$. }
  \label{fig:ht in su2}
\end{figure}

\begin{figure}[htbp]
	\centering
	\includegraphics[width=0.48\textwidth]{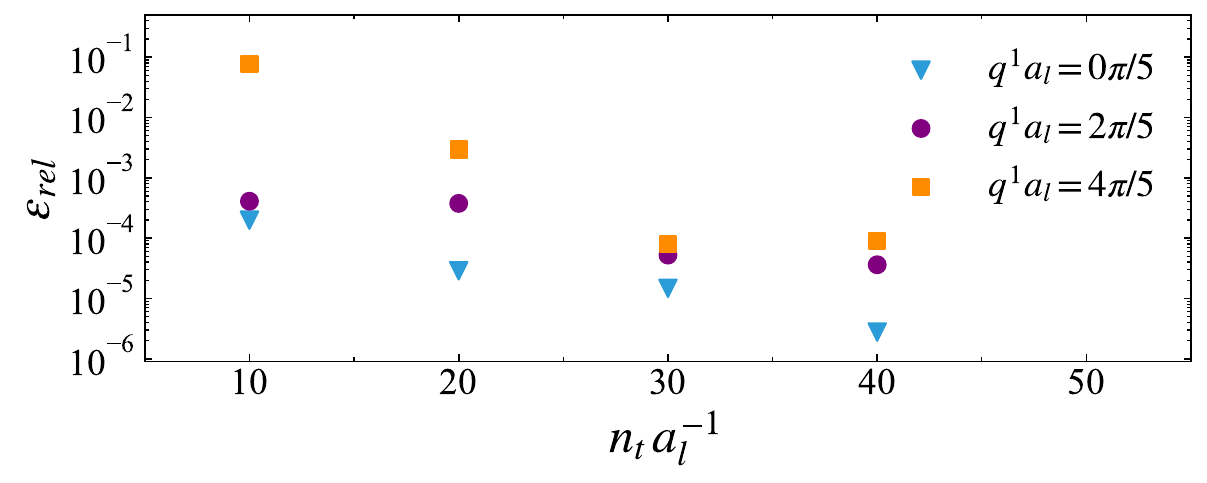}
	\caption{Relative errors of $|\mathcal{J}^0|^2$ from hadronic tensor with respect to direct calculation for time truncations $n_t$ in the $\rm SU(2)$ baryon case.
	}
	\label{fig:re_error_su2}
\end{figure}

The hadronic tensors of the lowest-lying $q\bar{q}$ meson and $qq$ baryon bound states are shown in Fig.~\ref{fig:meson_ht} and Fig.~\ref{fig:baryon_ht}, respectively. We also find a series of sharp peaks in $W^{00}(q)$ located in the exact hadron energy region, which, as discussed before, originate from one-particle final states $\ket{h'(p^\prime)}$. 
To isolate the contribution of each peak, integrate $W^{00}$ over a small region $R$ in $q^0$ that contains $p^{\prime 0}$ but excludes other peaks
\bea
    \int_R d q^0\, W^{00}(q) = \frac{2\pi}{2E_{p^\prime}} |\mathcal{J}^0|^2\,,
\eea
where $|\mathcal{J}^\mu|^2\equiv|\bra{h}J^\mu(0)\ket{h^\prime}|^2$.  
The region of integration $R$ must be chosen large enough to capture the broadened peak but small enough to exclude nearby peaks such that the resulting integral remains approximately invariant. 
For comparison, the matrix element $|\mathcal{J}^0|^2$ is also computed directly using exact diagonalization (ED).
Fig.~\ref{fig:re_error_su2} shows the relative errors $\varepsilon_{rel}$ between the hadronic tensor results $|\mathcal{J}_{HT}^0|^2$ and the direct calculation $|\mathcal{J}_{DC}^0|^2$ for various time truncations $n_t$, 
\bea
    \varepsilon_{rel}=\frac{|\,|\mathcal{J}_{HT}^0|^2-|\mathcal{J}_{DC}^0|^2\,|}{|\mathcal{J}_{DC}^0|^2}\,,
\eea
indicating that the chosen time truncation is sufficient to guarantee the required accuracy. 

\begin{figure}[htbp]
	\centering
	\includegraphics[width=0.44\textwidth]{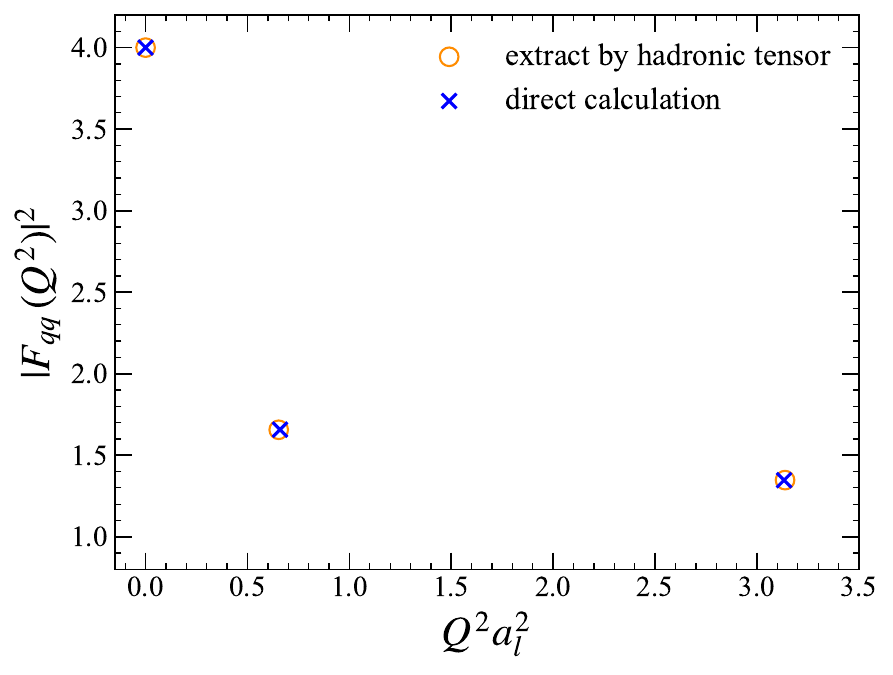}
	\caption{The $Q^2a^2$ dependence of $|F_{qq}(Q^2)|^2$ for the baryon.
	}
	\label{fig: Form factor of baryon}
\end{figure}

Also, the matrix element $\mathcal{J}^\mu$ can be expressed as the form factor through Lorentz symmetry:
\bea
    \bra{h(p')}J^{\mu}(0)\ket{h({p})}=(p+p^\prime)^{\mu} F(Q^2)\,.
\eea
In Fig.~\ref{fig: Form factor of baryon}, we compare the $|F_{qq}(Q^2)|^2$ obtained from direct calculation (yellow circles) and extracted from the hadronic tensor (blue crosses).
With unit fermion charge, the lowest-lying baryon carries total charge two, and its form factor square thus equals four at $Q^2a^2_l=0$. 
The baryon form factor exhibits a nontrivial $Q^2a^2_l$ dependence due to the internal structure of the state~\cite{Kubis:2000zd}.
The good agreement between the baryon form factor $|F_{qq}(Q^2)|^2$ obtained from direct calculation and that extracted from the hadronic tensor shows the correctness of our quantum algorithm as well as the results of the hadronic tensor in $\rm SU(2)$ LGT.

\section{summary}\label{sec:summary}

In this study, we have developed and validated a quantum algorithm for simulating the hadronic tensor in $\rm U(1)$ and $\rm SU(2)$ lattice gauge theories within the Hamiltonian formulation.
Using a quantum-number-resolving variational quantum eigensolver, both meson and baryon states can be prepared in well-defined symmetry sectors.
The hadronic tensor is then evaluated through real-time current--current correlation functions, which are accessed by introducing an auxiliary qubit.
We apply this algorithm to compute the hadronic tensor of the lightest meson in both $\rm U(1)$ and $\rm SU(2)$ lattice gauge theories, as well as that of the baryon in the $\rm SU(2)$ case, by simulating the quantum circuits on classical computers.
The resulting hadronic tensors exhibit clear spectral structures associated with one-particle intermediate states, in agreement with the Hamiltonian spectra.
We further demonstrate that charge-conjugation symmetry governs the selection rules of the electromagnetic current, leading to vanishing contributions from C-odd states while allowing nonzero contributions from C-even states.

Furthermore, we extract hadronic form factors from the hadronic tensor.
As a benchmark, the same form factors are independently calculated using exact diagonalization.
The good agreement between the two approaches validates both the quantum algorithm and the resulting hadronic tensors.

In the future, it is highly desirable to extend this framework to the direct extraction of parton distribution functions from the hadronic tensor.
Achieving this goal requires access to the deep inelastic region, where the factorization framework becomes applicable, and consequently demands much larger momentum transfers.
On the lattice, this translates into the need for finer momentum resolution and larger lattice sizes, which in turn require significantly more qubits.
Preliminary estimates suggest that accessing this kinematic region would require scaling the qubit count by a factor of two to three, surpassing the capabilities of current classical simulators.
This makes the hadronic tensor a promising observable for demonstrating the advantages of quantum computing in lattice gauge theory.

\begin{acknowledgments}
We thank QuNu members for valuable discussions. This work is supported by the National Natural Science Foundation of China with Project Nos. 12525508, 12475139, 12575085.
\end{acknowledgments}
%\newpage

%%%%%%%%%%%%%%%%%%%%%%%%%%%%%%%%%%%%%%%%%%%%%%%%%%%%%%%%%%%%
%%%%%%%%%%%%%%%%%%%%%%%%%%%%%%%%%%%%%%%%%%%%%%%%%%%%%%%%%%%%

%\bibliographystyle{h-physrev5}   
%\bibliography{refs.bib}

\begin{thebibliography}{10}

\bibitem{Collins:2011zzd}

\newblock J.~Collins{\em {Foundations of Perturbative QCD}} Vol.~32 (Cambridge
  University Press, 2011).

\bibitem{Liang:2019frk}
J.~Liang, T.~Draper, K.-F. Liu, A.~Rothkopf, and Y.-B. Yang, XQCD,
\newblock Phys. Rev. D {\bf 101}, 114503 (2020), arXiv:1906.05312.

\bibitem{Fang:2024ple}
Y.~Fang {\em et~al.},
\newblock Sci. China Phys. Mech. Astron. {\bf 68}, 260301 (2025),
  arXiv:2411.11294.

\bibitem{Zhang:2020uqo}
D.-B. Zhang, H.~Xing, H.~Yan, E.~Wang, and S.-L. Zhu,
\newblock Chin. Phys. B {\bf 30}, 020306 (2021), arXiv:2011.01431.

\bibitem{Bauer:2022hpo}
C.~W. Bauer {\em et~al.},
\newblock PRX Quantum {\bf 4}, 027001 (2023), arXiv:2204.03381.

\bibitem{Bauer:2023qgm}
C.~W. Bauer, Z.~Davoudi, N.~Klco, and M.~J. Savage,
\newblock Nature Rev. Phys. {\bf 5}, 420 (2023), arXiv:2404.06298.

\bibitem{Jordan:2011ci}
S.~P. Jordan, K.~S.~M. Lee, and J.~Preskill,
\newblock Quant. Inf. Comput. {\bf 14}, 1014 (2014), arXiv:1112.4833.

\bibitem{Li:2023kex}
T.~Li, W.~K. Lai, E.~Wang, and H.~Xing, QuNu,
\newblock Phys. Rev. D {\bf 109}, 036025 (2024), arXiv:2301.04179.

\bibitem{Lamm:2019uyc}
H.~Lamm, S.~Lawrence, and Y.~Yamauchi, NuQS,
\newblock Phys. Rev. Res. {\bf 2}, 013272 (2020), arXiv:1908.10439.

\bibitem{Li:2021kcs}
T.~Li {\em et~al.}, QuNu,
\newblock Phys. Rev. D {\bf 105}, L111502 (2022), arXiv:2106.03865.

\bibitem{Kang:2025xpz}
Z.-B. Kang, N.~Moran, P.~Nguyen, and W.~Qian,
\newblock (2025), arXiv:2501.09738.

\bibitem{Banuls:2025wiq}
M.~C. Ba\~nuls, K.~Cichy, C.~J.~D. Lin, and M.~Schneider,
\newblock (2025), arXiv:2504.07508.

\bibitem{Li:2022lyt}
T.~Li {\em et~al.}, QuNu,
\newblock Sci. China Phys. Mech. Astron. {\bf 66}, 281011 (2023),
  arXiv:2207.13258.

\bibitem{Li:2024nod}
T.~Li, H.~Xing, and D.-B. Zhang,
\newblock (2024), arXiv:2406.05683.

\bibitem{Barata:2025hgx}
J.~a. Barata and E.~Rico,
\newblock (2025), arXiv:2502.17558.

\bibitem{Ikeda:2025bjb}
K.~Ikeda, Z.-B. Kang, D.~E. Kharzeev, and W.~Qian,
\newblock (2025), arXiv:2512.18062.

\bibitem{Muschik:2016tws}
C.~Muschik {\em et~al.},
\newblock New J. Phys. {\bf 19}, 103020 (2017), arXiv:1612.08653.

\bibitem{Zache:2018jbt}
T.~V. Zache {\em et~al.},
\newblock Quantum Sci. Technol. {\bf 3}, 034010 (2018), arXiv:1802.06704.

\bibitem{Klco:2019evd}
N.~Klco, J.~R. Stryker, and M.~J. Savage,
\newblock Phys. Rev. D {\bf 101}, 074512 (2020), arXiv:1908.06935.

\bibitem{Calajo:2024qrc}
G.~Calaj{\`o} {\em et~al.},
\newblock PRX Quantum {\bf 5}, 040309 (2024), arXiv:2402.07987.

\bibitem{Atas:2021ext}
Y.~Y. Atas {\em et~al.},
\newblock Nature Commun. {\bf 12}, 6499 (2021), arXiv:2102.08920.

\bibitem{Zhang:2024fgv}
G.~Zhang, X.~Guo, E.~Wang, and H.~Xing,
\newblock Phys. Rev. D {\bf 111}, 056031 (2025), arXiv:2411.18869.

\bibitem{Lee:2024jnt}
K.~Lee, F.~Turro, and X.~Yao,
\newblock Phys. Rev. D {\bf 111}, 054514 (2025), arXiv:2409.13830.

\bibitem{Lowenstein:1971fc}
J.~H. Lowenstein and J.~A. Swieca,
\newblock Annals Phys. {\bf 68}, 172 (1971).

\bibitem{Abdalla:1991vua}
E.~Abdalla, M.~C.~B. Abdalla, and K.~D. Rothe,
\newblock {\em {Nonperturbative methods in two-dimensional quantum field
  theory}} (, 1991).

\bibitem{Coleman:1976uz}
S.~R. Coleman,
\newblock Annals Phys. {\bf 101}, 239 (1976).

\bibitem{Byrnes:2002nv}
T.~Byrnes, P.~Sriganesh, R.~J. Bursill, and C.~J. Hamer,
\newblock Phys. Rev. D {\bf 66}, 013002 (2002), arXiv:hep-lat/0202014.

\bibitem{Buyens2017fr}
B.~Buyens, S.~Montangero, J.~Haegeman, F.~Verstraete, and K.~Van~Acoleyen,
\newblock Phys. Rev. D {\bf 95}, 094509 (2017), arXiv:1702.08838.

\bibitem{Wilson:1974sk}
K.~G. Wilson,
\newblock Phys. Rev. D {\bf 10}, 2445 (1974).

\bibitem{Kogut1975hf}
J.~Kogut and L.~Susskind,
\newblock Phys. Rev. D {\bf 11}, 395 (1975).

\bibitem{Chen:2025zeh}
J.-W. Chen, Y.-T. Chen, and G.~Meher,
\newblock (2025), arXiv:2506.16829.

\bibitem{Hamer:1997dx}
C.~J. Hamer, W.-h. Zheng, and J.~Oitmaa,
\newblock Phys. Rev. D {\bf 56}, 55 (1997), arXiv:hep-lat/9701015.

\bibitem{Sala:2018dui}
P.~Sala {\em et~al.},
\newblock Phys. Rev. D {\bf 98}, 034505 (2018), arXiv:1805.05190.

\bibitem{Backens_2019}
S.~Backens, A.~Shnirman, and Y.~Makhlin,
\newblock Scientific Reports {\bf 9} (2019).

\bibitem{Nakanishi2019ss}
K.~M. Nakanishi, K.~Mitarai, and K.~Fujii,
\newblock Phys. Rev. Res. {\bf 1}, 033062 (2019).

\bibitem{Bartschi:2019zce}
A.~B\"artschi and S.~Eidenbenz,
\newblock Lect. Notes Comput. Sci. {\bf 11651}, 126 (2019).

\bibitem{Pedernales:2014izf}
S.~Pedernales, J.~\, R.~D. Candia, L.~Egusquiza, I.~\, J.~Casanova, and
  E.~Solano,
\newblock Phys. Rev. Lett. {\bf 113}, 020505 (2014).

\bibitem{nielsen_chuang_2010}
M.~A. Nielsen and I.~L. Chuang,
\newblock {\em Quantum Computation and Quantum Information: 10th Anniversary
  Edition} (Cambridge University Press, 2010).

\bibitem{Phillip2017qa}
P.~Weinberg and M.~Bukov,
\newblock SciPost Phys. {\bf 2}, 003 (2017).

\bibitem{Harris:1978yaq}
F.~J. Harris,
\newblock 1978.

\bibitem{White:2004etd}
S.~R. White and A.~E. Feiguin,
\newblock Phys. Rev. Lett. {\bf 93}, 076401 (2004), arXiv:cond-mat/0403310.

\bibitem{Kubis:2000zd}
B.~Kubis and U.-G. Meissner,
\newblock Nucl. Phys. A {\bf 679}, 698 (2001), arXiv:hep-ph/0007056.


\end{thebibliography}

%%%%%%%%%%%%%%%%%%%%%%%%%%%%%%%%%%%%%%%%
\end{document}